# Post-Quantum Blockchain: Challenges and Opportunities

Sufyan Al-Janabi

College of Computer Science and IT, University of Anbar, Ramadi, Iraq

sufyan.aljanabi@uoanbar.edu.iq, https://orcid.org/0000-0002-2805-5738

**Abstract:**

*Blockchain is a Distributed Ledger Technology (DLT) that offers numerous benefits including decentralization, transparency, efficiency, and reduced costs. Hence, blockchain has been included in many fields. Blockchain relies on cryptographic protocols (especially public-key cryptography and hash functions) to achieve many essential sub-routines. However, the increased progress of quantum computation and algorithms has threatened the security of many traditional cryptosystems. Therefore, this represents a serious risk for the existing blockchain technology. For example, SHA-256 and the Elliptic Curve Digital Signature Algorithm (ECDSA) cryptosystems can be compromised by Shor's and Grover's quantum algorithms in the foreseeable future. Post-Quantum Cryptography (PQC) is a basic solution for resisting these quantum attacks. Applying PQC to blockchains results in creating Post-Quantum Blockchains (PQB). Thus, this paper aims to review the threats imposed by quantum computers on "classical" blockchain technology and provide useful guidelines on PQB security to blockchain researchers. The paper focuses on the challenges and opportunities of future work direction in this field.*

**Keywords:** Blockchain, Cybersecurity, Post-quantum cryptography, Quantum-resistant, Quantum computing.

## 1. Introduction

Blockchain technology provides a distributed Peer-to-Peer (P2P) facility for value transactions without the mediation of a central authority. Blockchain originated as the underlying mechanism of the famous peer-to-peer Bitcoin cryptocurrency. Based on its ability to use consensus mechanisms in a trustless environment, blockchain can shortly change many of our life domains. On the one hand, cloud computing as a centralized approach provides some advantages including affordability and resource availability. However, cloud computing has an important security concern of single-point failure. On the other hand, blockchain can potentially address these security concerns by providing an immutable and decentralized shared ledger for maintaining asset and transaction records over P2P networks [1].

Blockchain technology is capable of providing features like secure communications, transparency, data privacy, and resilience. This has popularized blockchain in recent years and thus it has been proposed as a key technology for various applications other than cryptocurrencies including smart health, e-voting, logistics, measuring systems, IoT, and smart factories.

Typically, cryptosystems aim to protect an information asset. One peculiar property of blockchain that differentiates it from other cryptosystems is that a blockchain is a ledger. Hence, a blockchain is the asset itself. Blockchain security is achieved using cryptographic techniques. In particular, public-key schemes such as RSA (which relies on the difficulty of large integer factorization) or Elliptic Curve (EC) cryptosystems (which rely on the discrete logarithm problem) are used for the generation of public/private key pairs needed to protect the data assets stored on blockchains. In traditional security systems, the data is decoupled from the key pairs. Thus, when a cryptographic key is compromised, a central authority can be used to revoke its validity. Thereafter, the central authority can issue a new key pair and associate it with the data. In contradiction, a blockchain has no central authority for managing users' keys. It is assumed that the resource owner is the one keeping the private keys. Hence, when a key is compromised, the data asset might be irrevocably stolen. This means that it is not easy to decouple the protected resources from the cryptosystem being used in a blockchain. For this particular reason, blockchains are impacted by the advancements in quantum computing.

Blockchain technology is continuing to get much interest from various sectors such as government, banking, insurance, healthcare, and transportation because it can offer ownership verification, transparency, security, and privacy. The current blockchain data structures offer these benefits based on public-key cryptography and hash functions. However, emerging quantum technologies can have a huge impact on nowadays, especially those based on computationally hard problems. Quantum computation is based on exploiting some quantum physical effects in order to decrease the time requirement for solving certain computational problems. This is mainly achieved by the creation and utilization of quantum superpositions. Quantum computers can run Shor's algorithm to exploit cryptographic systems that are based on discrete logarithmic problems, integer factorization, and finding the hidden subgroup in abelian finite groups. Indeed, another quantum procedure based on Grover's algorithm can achieve a quadratic speed in searching in unordered collections [2].

A powerful quantum computer can use Shor's algorithm to break some popular public-key algorithms in polynomial time. This includes RSA, the Elliptic Curve Digital Signature Algorithm (ECDSA), the Elliptic Curve Diffie-Hellman (ECDH), and the Digital Signature Algorithm (DSA). Furthermore, quantum computers can use Grover's algorithm for accelerating the generation of hashes. This might enable the recreation of the entire blockchain. Moreover, Grover's algorithm can be used to detect hash collisions in order to replace blocks of the blockchain while preserving its integrity. Therefore, it is significantly important to thoroughly understand the current level of vulnerability of blockchain systems to quantum attacks. This is due to the strong embedded coupling between data and cryptosystems used in blockchains, the potential vulnerability of the famous cryptosystems to quantum computer attacks, and the possible introduction of sufficiently powerful quantum computers in the mid-range future [3].

This paper reviews the current blockchain technology and the traditional cryptosystems used for achieving blockchain security. It also studies the security impact of quantum attacks on this technology. The main Post-Quantum Cryptography (PQC) schemes are discussed and their use for the creation of Post-Quantum Blockchains (PQB) is outlined. This enables us to track the main future research direction in the field. The remainder of this paper is organized as follows. Section 2 gives a short overview of blockchain technology before discussing its vulnerability to quantum attacks in Section 3. Next, the main PQC schemes are presented in Section 4 with an emphasis on the NIST PQC adopted standards. Section 5 outlines the most important available PQB schemes. Then, the main research challenges and opportunities are discussed in Section 6. Finally, Section 7 concludes the paper.

## 2. Blockchain Overview

Blockchain is a decentralized and distributed ledger that offers the following main security features [3], [4]:

- Data immutability: This means once a transaction is added to the blockchain, it cannot be altered or deleted. A transaction can only be added after approving its validity by the majority of blockchain nodes.
- Decentralization and distributivity: Each blockchain node has access to the entire distributed ledger. This is controlled using the consensus algorithm. Thus, if one node is compromised or shut down, other nodes of the blockchain still have its information.
- Integrity and data privacy: This is mainly achieved in the blockchain using public-key cryptosystems and hash functions.
- Reduced costs and efficiency: The cost is by using a single shared and distributed ledger. This enables real-time auditing and settlement by all parties whenever a transaction occurs.

Therefore, blockchain technology has many applications such as finance, insurance, healthcare, the Internet of Things (IoT), e-voting, and supply chain [1], [5]. According to the roles of different nodes in the system, it is possible to mention the following types of blockchains [4]:

- Permissionless and fully decentralized public blockchains (Examples include Bitcoin, Monero, Ethereum, Litecoin, and Zerocash).
- Permissioned and partially decentralized public blockchains (Such as Ripple, Libra, and EOS).
- Permissionless private blockchains (LTO is an example of this type).
- Permissioned private blockchains (Examples include Multichain and Monax).
- Federated blockchains or consortium blockchains that combine the permissioned public blockchain and the permissionless private blockchain types (Such as Hyperledger, Corda, and Quorum).
- Hybrid blockchains that integrate permissioned private and permissionless public systems.

It is possible to simplify blockchain technologies into two constituent parts. These are the transaction mechanism and the consensus protocol. The transaction mechanism decides the method of transferring tokens and information in the blockchain. This requires using digital signatures to authenticate the possession of the used public and private keys. The consensus protocols enable the establishment of agreement among the blockchain nodes to ensure that the distributed information is consistent and accurate [6]. There are a variety of consensus mechanisms, each one has its pros and cons. However, the most famous two of them are:

1. Proof of Work (PoW): The miners try to guess the answer to a certain mathematical problem. Whenever a miner node guesses, it will construct a block that can be accepted as a legitimate block if and only if it is the one with the most accumulated PoW.
2. Proof of Stake (PoS): A new block is added to the blockchain based on a procedure influenced by the number of coins staked in the blockchain network.

Blockchain systems have to be secure and well-protected against various threats such as tampering, fraud, and unauthorized access. The most important cryptographic techniques necessary to enhance blockchain security are [3], [7]:

- Digital signatures: Digital signatures are crucial to demonstrate that a transaction in a blockchain has been approved by the legitimate holder of the private key. This demonstrates the authorization and validity of transactions.
- Hash Functions: Hash functions are commonly used cryptographic primitives in blockchains. They are used in generating digital signatures and also to link blockchain' blocks. This chaining ensures the blockchain's immutability.
- Merkle trees: Merkle trees enable the arrangement and compilation of the transactions in a block. Hence, the participants can easily confirm the existence and integrity of a transaction in a block without processing all other individual transactions.
- Public key infrastructure (PKI): Blockchain's participants use PKI to validate each other's public keys. Thus, they can confirm the validity of transactions.

Besides these primary security primitives of the blockchain, there are some other optional primitives. These include special signatures, zero-knowledge proof, secret sharing, accumulators, and commitments. The use of these optional primitives can improve the privacy, traceability, and anonymity of the transactions.

## 3. Blockchain Vulnerability to Quantum Attacks

The main quantum computation concepts were shaped in the 1980s by Paul Benioff, Yuri Manin, and Richard Feynman. The difficulty for classical computers to simulate quantum systems that evolve in time was noted. Indeed, it was postulated that it is possible to build machines that work under the laws of quantum physics to simulate quantum systems efficiently. From the point of view of blockchain vulnerability, there are two main types of quantum algorithms. These are subgroup-finding algorithms and amplitude amplification algorithms [4].

For the first category, Simon's was the first algorithm proven to achieve exponential speedup on quantum computers. This algorithm inspired one of the main breakthroughs in quantum computing which is Shor's algorithm that can solve large integers' factorization and discrete logarithm problems in polynomial time. This is particularly important because most currently deployed public-key cryptosystems such as RSA, Diffie-Hellman, EC, and ElGamal are based on the hardness of one of these two computational problems. This can be a crucial issue in blockchain security because of the possibility of compromising the authenticity of digital signatures. Moreover, it is possible to efficiently solve the Hidden Subgroup Problem (HSP) over finite abelian groups in polynomial time using generalizations from Simon's and Shor's algorithms.

Taking the breaking RSA-2048 case as an example, a 5 GHz CPU classical computer can roughly need 13.7 billion years to do this using the best currently known techniques. On the other hand, a 10 MHz quantum computer would be able to achieve this task in roughly 42 minutes. However, such a quantum computer needs quantum memory that is at least large enough to hold a state that represents both the input and the output [2].

The second category of algorithms (amplitude amplification) is based on generalizations of Grover's algorithm that represent another breakthrough in quantum computing. Grover's algorithm enhances the search in unordered collections of size N allowing for NP-complete problems to be solved on quantum computers quadratically faster than on any classical device. Despite the speed-up here is not as dramatic as in the first category (it is possible to mitigate this attack by doubling the size of the key), the importance of the algorithms of the second category relies upon their general applicability. The algorithms of amplitude amplification are particularly important because most blockchain technologies use consensus algorithms that are based on solving NP-complete problems. Quantum collision search and quantum counting are two interesting applications inspired by Grover's algorithm.

Quantum computers are still now in their early stages. One of the most powerful quantum computers currently available is the IBM Osprey has 433 qubits only, while breaking RSA-2048 would require a 4000-qubit quantum machine. Nevertheless, it is no longer possible to exclude the deployment of quantum-safe cryptosystems. This is because an adversary can follow a strategy for collecting some important ciphertexts encrypted by the available traditional cryptosystems, storing them, and then decrypting them when powerful quantum computers become available [8]. Therefore, developing new cryptosystems that can withstand future quantum attacks has become a hot research topic. There are two main directions in this area, which are the solutions based on quantum cryptography and those based on PQC. Quantum cryptography explores the principles of quantum mechanics for performing secure cryptography tasks and detecting the existence of eavesdroppers. Here, the Quantum Key Distribution (QKD) is the most established quantum protocol. However, quantum cryptography solutions cannot be directly used over the existing security infrastructure [9]. Therefore, this paper is mainly focusing on PQC-based solutions. PQC constructions are classical cryptosystems based on some NP-hard problems that are implemented on classical computers but can resist quantum computers-based attacks.

It is not possible to give accurate predictions of the progress of future technology. However, we can do an extrapolation of the past and current trends in the advancement of quantum technology including the fidelity of gates, number of qubits, fault-tolerance, and error correction. Thus, some well-established researchers share the conclusion that it is quite likely that quantum computers capable of breaking RSA-2048 can be built by the year 2035. Regardless of the accuracy of such extrapolation, people have to continue building communication channels with higher security that might result in a cryptographic revolution. Thus, the US National Institute of Standards and Technology (NIST) has started the standardizing process of quantum-resistant public-key cryptosystems [7].

Typically, the security level is measured by the amount of computation needed by a machine to implement a brute-force attack. Nonetheless, the NIST uses estimates of classical and/or quantum gate counts required for attack to provide broader security categories that range from 1 to 5. The vulnerabilities of some key blockchain technologies have been studied by many researchers. For example, Table 1 shows the vulnerabilities of some famous cryptocurrencies (Bitcoin, Ethereum, Litecoin, Monero, and ZCash) against the two main forms of quantum attacks (Shor's and Grover's). All blockchains have strong vulnerabilities against Shor's algorithm attacks because of the exponential advantage it offers. However, most cryptocurrencies have an intermediate level of vulnerability against Grover's algorithm because of the quadratic quantum advantage it achieves. Of the mentioned blockchains, Monero can be considered the most secure against quantum attacks due to its procedure for the obfuscation of transacted values [2].

Table 1: Vulnerabilities of some key blockchain technologies to quantum attacks [2].

|   | Blockchain (cryptocurrency) | Vulnerability to subgroup-finding (Shor's) attack | Vulnerability to amplitude amplification (Grover's) attack |
|---|---|---|---|
| 1 | Bitcoin | strong | intermediate |
| 2 | Ethereum | strong | intermediate |
| 3 | Litecoin | strong | intermediate |
| 4 | Monero | strong | currently considered safe |
| 5 | ZCash | strong | intermediate |

## 4. Post-Quantum Cryptography

The main objective of PQC is to build cryptosystems that can withstand attacks from both quantum and classical computers while being able to use the existing network infrastructures. Thereafter, PQC has gained noticeable research and academic attention. Hence, some interesting projects (e.g. PQCrypto) have been established for educating researchers and developers about this field. Furthermore, in 2016 the NIST initiated a process to adopt new PQC standards for both digital signatures and key establishment. This process has resulted in the announcement of four PQC standards so far as will be described later.

Most of the current "pre-quantum" cryptosystems rely on integer factorization or discrete logarithm problems. However, such schemes will be obsolete when powerful quantum computers are introduced. In contrast, the PQC counterparts are based on hardness problems that are believed to retain their hardness even with the introduction of quantum computers. The main features that are needed by PQC schemes to efficiently support PQB are small key sizes, small hash and signature length, low computational complexity, fast execution, and low energy consumption [10]. The five main types of PQC hardness assumptions are summarized in the following subsection.

### 4.1 PQC Types

The main classes of PQC are the following [3], [4], [8]:

- *Lattice-based cryptography:* The Shortest Vector Problem (SVP) is a hard cryptographic problem that relies on the security of lattice-based cryptosystems. It can also be reduced to another two hard problems, which are the Shortest Basis Problem and Closest Vector Problem. Learning With Errors (LWE) is a special class of mathematically hard problems particularly related to lattice-based cryptography. Lattice-based schemes have a good balance in terms of key size, security level, and computational simplicity. Therefore, the NIST has accepted the Kyber, FALCON, and Dilithium lattice-based schemes as standards for key encapsulation and digital signature.
- *Code-based cryptography:* Code-based schemes can be the oldest type of PQC. They are based on the concept of error-correcting codes. The basic McEliece scheme, for example, introduced in 1978 has a structure that made it resistant to quantum computers. The high-security level of this type of cryptosystem has been justified by thorough analysis during the last four decades. Unfortunately, the disadvantage of code-based schemes is their significant key size. Hence, it is challenging to implement them on resource-limited devices. Thus, there have been various proposals to reduce the key length of these cryptosystems.
- *Hash-based cryptography:* Hash-based cryptosystems are a specific type of PQC used for digital signatures only. Their security depends on the security of the underlying hash function. There are variants of hash-based schemes that are derived from the Merkle tree that are considered now as promising hash-based signature schemes for PQC. Hash-based schemes can be divided into two main classes, which are stateless and stateful schemes. SPHINCS+ chosen by the NIST as a standard for digital signatures is based on a stateless approach.
- *Multivariate-based cryptography:* This PQC category is founded on the complexity of solving a certain type of polynomial equations that are known as multivariate equations. These problems are difficult to solve in finite fields. This is why they are considered resistant to quantum attacks. The type of multivariate-based scheme depends on the degree of the polynomial equation (e.g., quadratic, cubic, etc.).
- *Supersingular elliptic curve isogeny cryptography:* This type of cryptosystem is the youngest PQC algorithm. They are founded on isogeny as a new kind of ECC. In general, these schemes are divided into three main classes, which are: Ordinary Isogeny Diffie-Hellman (OIDH), Supersingular Isogeny Diffie-Hellman (SIDH), and Commutative SIDH (CSIDH).

## 4.2 NIST PQC Standards

The NIST has already endorsed four PQC systems. The key-establishment scheme that has been selected for standardization is CRYSTALS‑KYBER. The three digital signature schemes that have been approved are CRYSTALS‑Dilithium, SPHINCS+, and FALCON. However, CRYSTALS‑Dilithium is recommended by the NIST as the primary algorithm for implementation. Indeed, four other key establishment candidates will advance to the fourth evaluation round for future standardization. These are BIKE, HQC, Classic McEliece, and SIKE. In this subsection, the NIST-nominated four PQC schemes are overviewed.

The key establishment protocol is used to securely enable two parties to share the same secret (symmetric) key. This is done by an asymmetric crypto-mechanism known as Key Encapsulation Mechanism (KEM). The established secret key can then be used for message encryption. Thus, KEM is a kind of key exchange protocol. The necessity of using PQC to exchange symmetric keys has become clear due to the recent advances in quantum computing. NIST recently selected the CRYSTALS-Kyber encapsulation algorithm as a PQC key establishment standard. This scheme enables two parties to establish confidentiality and trust without being compromised by an Indistinguishable under non-adaptive and Adaptive Chosen Ciphertext Attack (IND-CCA1/IND-CCA2). The fundamentals of the CRYSTALS-Kyber scheme are founded on Module-LWE (a variation of the LWE problems). This is supposed to be an NP-hard lattice problem. The scheme supports three various security levels, each with a unique set of parameters. Approximately, the security levels achieved by Kyber-512, Kyber-768, and Kyber-1024 are 128 (1), 192 (3), and 256 (5) bits, respectively [11].

On the other hand, digital signatures have extensive uses in nowadays security systems. For example, blockchain applications heavily rely on digital signatures for transaction validation and for showing the legitimacy of nodes in the network. Similarly to pre-quantum signature schemes, the security assumption for a PQC signature implies that an adversary with access to a signing oracle cannot add a new signature to a previously signed message nor change the signature of an already-signed message. In fact, some PQC signatures are already used in some applications based on earlier NIST recommendations. In addition, NIST has recently selected Dilithium, FALCON, and SPHINCS+ as three PQC signature schemes as future deployment standards.

CRYSTALS-Dilithium is a PQC signature that is founded on the hardness of module lattice problems. Although Dilithium's signature is larger than some other lattice-based signature schemes, it achieves a much lower public key size in comparison to them. The dilithium scheme is developed based on the Fiat-Shamir Abort framework. It uses standard Number Theoretic Transform (NTT) polynomial multiplication with integer instructions in its vectorized version. It utilizes matrix and vectors that are driven by SHAKE128 and SHAKE256. Thus, matrix and vector expansion is accelerated by its vectorization.

The second PQC standard scheme is SPHINCS+, which is a stateless hash-based algorithm. This scheme has been designed with three hash functions, which are SHA-256, SHAKE256, and Haraka. It produces a large signature size of about 49 KB. However, it has the best private and public key sizes of less than 1 KB depending on the selected parameters. SPHINCS+ effectively employs a classical Merkle hash tree called the Many-Tree Signature (MTS). The scheme utilizes a hypertree linked using One-Time Signatures (OTS) for the authentication of many Few-Time Signature (FTS) key pairs.

At last, FALCON is a lattice-based PQC signature algorithm that has been chosen as the standard. It utilizes a trapdoor sampler known as rapid Fourier sampling for establishing the framework over NTRU lattices. FALCON employs the Short Integer Solution (SIS) hard problem over NTRU. This scheme offers relatively small private and public key sizes. It also produces very small signature sizes. Indeed, FALCON has a fast and relatively easy verification step. However, it has a clear disadvantage in terms of the time requirement for key generation compared to the other two standard schemes.

In summary, Dilithium produces significantly larger sizes of public and private keys than SPHINC+. However, Dilithium provides a much better digital signature size and speed compared to FALCON and SPHINC+. Nevertheless, FALCON and SPHINC+ are more suitable for limited memory devices because they outperform Dilithium in terms of memory usage. A comparison study evaluated some PQC signature techniques with the ECDSA using an implementation based on the Open Quantum Safe (OQS), as shown in Table 2. The table estimates the time required by each signature algorithm to implement hashing and LibOQS functions for each block in comparison to the algorithm's public key plus signature size. The calculated time includes hashing times, key generation, signature generation, and signature verification [1].

Table 2: Public keys and signature sizes of some signature algorithms [1].

|   | Algorithm | Size of public key and signature (bytes) | Execution Time (ms) |
|---|---|---|---|
| 1 | ECDSA | 96 | 4 |
| 2 | Falcon 512 | 1563 | 18 |
| 3 | Falcon 1024 | 3073 | 28 |
| 4 | Dilithium 2 | 3228 | 18 |
| 5 | Dilithium 3 | 4173 | 21 |
| 6 | Dilithium 4 | 5126 | 25 |

## 5. Post-Quantum Blockchain Schemes

Blockchains are impacted by the advancement of quantum computers similarly to many other security protocols and applications relying on traditional "pre-quantum" cryptography. Hence, most well-known blockchains are currently considering PQC for their future development stage. For example, a branch of Bitcoin's blockchain is experimenting with quantum-resistant mining and signatures. This is expected to lead to the development of post-quantum Bitcoin. Another example is Ethereum 3.0, which aims to improve the security and scalability of Ethereum by utilizing quantum-safe elements such as Zero-Knowledge Scalable Transparent ARguments of Knowledge (zk-STARKs) [8].

Algorand has also introduced the usage of lattice-based post-quantum signatures. Indeed, other blockchains including IoTeX and Cardano are preparing their platforms for the era of quantum computing. Moreover, several PQB blockchain platforms (such as Nexus, Komodo, Tidecoin, QRL, and QANplatform) are currently operational. Each of these has its own purpose. Different researchers worldwide are working on PQB proposals or on modifications of traditional blockchains to react to quantum threats. The works on developing PQBs typically focus on three main aspects, which are: proposing quantum-safe signature algorithms, providing quantum-safe mining mechanisms, and securing communications in blockchain networks [10].

The authors in [12] presented PQFabric, which is an implementation of Hyperledger Fabric facilitated with quantum-resistant signatures. This implementation offers crypto-agility by performing fast migration to a hybrid quantum-resistant blockchain and choosing any existing signature algorithm from the Open Quantum Safe (OQS) library for each node. However, the work highlighted a hashing bottleneck due to the significant increase in hashing time. One possible future solution for this issue is to find a method for efficiently and securely encoding the public keys and signatures. This might result in significant speedups for PQC algorithms having large public keys and signatures.

The authors in [13] proposed a blockchain system based on lattice-based cryptography for resisting quantum attacks. The challenge was that the sizes of public keys and signatures required by lattice ciphers are much larger than those of traditional cryptosystems. This would degrade the speed and performance of the PQB because only a small number of transactions can be accommodated in each block. To deal with this problem, they suggested a method to only store the hash values of public keys and signatures in the PQB, while storing the complete versions of them in an Interplanetary File System (IPFS). Hence, the number of bytes needed by each transaction could be greatly reduced.

Another PQB work based on Hyperledger Fabric was introduced in [14]. The open-source Hyperledger Fabric framework founded by the Linux Foundation seems to be interesting for many PQB researchers due to its modularity, versatility, and robustness for enterprise blockchain platforms. The work provided an extension of the Hyperledger Fabric. While trying to maintain the main structure of the framework unchanged, the PQCrypto library was used to integrate the PQC schemes into Golang (the language of Fabric). This facilitated invoking the NIST algorithms. The changes made to the source code of Hyperledger Fabric mainly focused on creating data structures and functions necessary to invoke the PQC algorithms. The work dealt with PQC digital signatures. A future enhancement of the work could be to further extend using PQC in the code, e.g. by using KYBER or other algorithms for KEM.

The work in [15] considered the development and implementation of PQB within the IoT environment. The research introduced a PQB version based on Hyperledger Fabric that leverages quantum-safe signatures using the OQS library. The PQB was tested against NIST PQC round 3 candidates for measuring the optimal performance. The work generally demonstrated the potential of PQB technology for IoT applications. It focused on a specific blockchain platform (Hyperledger Fabric) and specific PQC algorithms. Thus, it could be extended by considering other blockchain platforms and other PQC schemes. Indeed, future empirical research is needed for evaluating the scalability, adaptability, and practicality of such proposals in real-world, large-scale IoT systems.

In overall, it is prudent to conclude that it is not straightforward for developers to choose a certain PQB cryptosystem. Such a decision will be affected by several parameters. These include the required security level, the blockchain node hardware, the necessary performance of the blockchain node, and the available resources (e.g., processor, memory, energy). The PQB field is still in its infancy and more thorough research is required to investigate the various aspects of it.

## 6. Challenges and Future Research Directions

This section is dedicated to investigating some of the fundamental research challenges and opportunities related to the development of PQBs. These could be related to the difficulties of deploying PQC schemes or concerned with the characteristics of blockchain. Thus, this section gives helpful insights for future research directions in the field as follows.

- *Evolution of quantum computing:* Quantum computing is now an active topic that is attracting major attention from both academia and industry. Therefore, PQB researchers should continuously pay attention to the possibility of the emergence of new quantum attacks against traditional and/or PQC cryptosystems.

- *Large size of keys and signatures:* PQC algorithms require significantly much larger key and signature sizes compared to traditional cryptosystems. This can be identified as a crucial disadvantage of PQC schemes that discourages their adoption. Hence, more research is required to achieve good trade-offs between security and key sizes of PQB technology.

- *Computational cost and energy consumption:* PQC schemes require more storage and computational processes. This would increase energy consumption, which may represent a serious challenge for resource-constrained devices. Indeed, some consensus protocols (such as PoW) usually consume a lot of energy. Thus, it is necessary to find new approaches for the optimization of PQC cryptosystems to maximize their energy and computational efficiency. Thereafter, the efficiency of PQB can be maximized.

- *Authentication and access management:* Traditionally, authentication and authorization tasks are done based on using trusted centralized authorities. On the other hand, some distributed approaches use blockchain smart contracts for authentication and identity management in large-scale ecosystems. However, these smart contract-based approaches are vulnerable to quantum attacks. Hence, there is a need for innovative approaches that utilize PQBs for authentication and access management in large-scale environments such as IoT.

- *Large execution time and ciphertext overheads:* Certain PQC cryptosystems need a significant execution time and generate ciphertext large overheads this can may impact the efficiency and performance of PQB. Therefore, finding optimum methods to reduce the execution time in PQC implementations and considering innovative compression techniques to minimize their ciphertext overhead are welcomed future research directions.

- *Security threats:* PQC schemes are designed to resist attacks from quantum computers. Nonetheless, they might still be vulnerable to other kinds of security threats. Some security threats that need to be also considered in the PQB setting include double spending, side-channel attacks, Sybil attacks, DDoS attacks, privacy concerns, hybrid attacks, transaction malleability, and smart contract vulnerabilities. This emphasizes the necessity of considering all relevant security concerns in any PQB application and using various techniques for mitigating these threats.

- *Real-time applications:* in traditional blockchains, consensus mechanisms confirm the transactions' validity within a predefined time. This is necessary to prevent forking in the blockchain. Moreover, some real-time applications imply a certain latency to ensure seamless and smooth network communications. Given the execution time challenge of PQC, there is a clear potential to design adaptive consensus mechanisms that achieve a good trade-off between maintaining decentralization and latency.

- *Quantum cryptography-based blockchain:* The QKD protocol offers guaranteed secrecy against potential eavesdropping even in the case that the eavesdropper has a quantum computer. Thus, there is a research direction to build quantum-safe blockchains based on quantum cryptography (For more details on this issue, the reader is advised to refer to [16]). Despite the technology challenge of this approach, future research might result in a breakthrough in quantum-safe blockchain technology. To the best of our knowledge, there are no major research initiatives to integrate PQB technology with quantum cryptography implementations. The pros and cons of such integration are not clear yet.

- *Hybrid cryptosystems:* These cryptosystems merge "pre-quantum" and post-quantum cryptography for achieving data security against quantum attacks and other security attacks against PQC schemes. This type of system has been considered by industry and academia. Such schemes are interesting. However, they need high computational resources because they include the implementation of two kinds of complex cryptosystems. Hence, future research is required to investigate the effectiveness of these cryptosystems and to reach a trade-off among security level, computational complexity, and energy consumption.

- *Additional security features:* It is possible to enhance the PQB security by leveraging certain security constructions and protocols that are not so far widely incorporated in commercial blockchain developments. These constructions include ring signatures, aggregate signatures, blind signatures, Identity-Based Encryption (IBE), homomorphic encryption, Secure Multi-Party Computation (SMPC), and zero-knowledge proofs. However, the security level and efficiency of these cryptographic constructions need more validation in the post-quantum era.

- *Consensus protocols:* The development of consensus mechanisms for the PQB is another hot research topic. The research covers two major directions. The first involves the development of novel posy-quantum consensus protocols. The second direction deals with the analysis and improvement of the conventional consensus protocols to efficiently withstand quantum attacks. More studies are necessary to analyze post-quantum solutions for consensus protocols. One important issue here is related to PoW and Grover algorithm-based attacks. Even though PoW is the most commonly deployed consensus mechanism in traditional blockchains, PoW needs a computational problem that is efficient for verification. This implies NP computational problems, which are amenable to Grover's algorithm attack. Hence, any PoW system is inherently susceptible to Grover speed-up. This in turn means that a miner with a quantum computer will always have an advantage over a miner with a classical one in all blockchains that utilizes the PoW mechanism. To counter this, future PQB developers need to consider alternative consensus mechanisms such as PoS.

- *InterPlanetary File System (IPFS):* Some previous studies proposed the integration of PQC signatures with the IPFS to enhance the PQB efficiency. This can be done by only storing the hash values of public keys and digital signatures in the PQB while storing their complete content in the IPFS. The implication of integrating the PQC and IPFS on the long-term dependability and viability of PQB still needs more investigation. More important is to analyze the security and complexity of the IPFS components and the whole integrated system.

# 7. Conclusion

The fast advancements of quantum computing represent a serious threat to many traditional cryptosystems. Thus, in order to make the most of blockchain potential, it is necessary to maintain its security against quantum attacks. This article considered the impact of attacks by quantum computers (mainly based on Shor's and Grover's algorithms) on the current blockchain technology. One major research direction to avoid the threats of quantum technology is to use PQC schemes for building PQBs. Both blockchain and quantum computing technologies can still be considered as a fledgling technology. Hence, there is an available time for adaptation and course correction. We have outlined the major future research challenges and opportunities in the field so that future blockchain developers can tune their focus on the main issues and concerns of the PQB.

# References:

[1] Lu Gan and Bakhtiyor Yokubov, "A performance comparison of post-quantum algorithms in blockchain," *The JBBA*, Vol. 6, No. 1, 2023, pp. 1-10, doi: 10.31585/jbba-6-1-(1)2023

[2] Joseph J. Kearney and Carlos A. Perez-Delgado, "Vulnerability of blockchain technologies to quantum attacks," *Array*, Vol. 10, 100065, 2021, doi: 10.1016/j.array.2021.100065.

[3] T. M. Fernández-Caramès and P. Fraga-Lamas, "Towards Post-Quantum Blockchain: A Review on Blockchain Cryptography Resistant to Quantum Computing Attacks," *IEEE Access*, Vol. 8, pp. 21091-21116, 2020, doi: 10.1109/ACCESS.2020.2968985.

[4] Andrada-Teodora Ciulei, Marian-Codrin Crețu, and Emil Simion, "Preparation for Post-Quantum era: a survey about blockchain schemes from a post-quantum perspective," *Cryptology ePrint Archive*, Paper 2022/026, 2022, https://ia.cr/2022/026

[5] Saba Salman, Sufyan Al-Janabi, and Ali M. Sagheer, "Valid Blockchain-Based E-Voting Using Elliptic Curve and Homomorphic Encryption," *International Journal of Interactive Mobile Technologies*, Vol.16. No. 20, 2022, pp. 79-97.

[6] J. Gomes, S. Khan and D. Svetinovic, "Fortifying the Blockchain: A Systematic Review and Classification of Post-Quantum Consensus Solutions for Enhanced Security and Resilience," *IEEE Access*, Vol. 11, pp. 74088-74100, 2023, doi: 10.1109/ACCESS.2023.3296559.

[7] P. Thanalakshmi, A. Rishikesh, Joel Marion Marceline, Gyanendra Prasad Joshi, and Woong Cho, "A Quantum-Resistant Blockchain System: A Comparative Analysis," *Mathematics*, Vol. 11, No. 3947, 2023, doi: 10.3390/math11183947

[8] H. Gharavi, J. Granjal, and E. Monteiro, "Post-Quantum Blockchain Security for the Internet of Things: Survey and Research Directions," *IEEE Communications Surveys & Tutorials* (Early Access), Jan. 2024, doi: 10.1109/COMST.2024.3355222.

[9] Sufyan T. Faraj, "A Novel Extension of SSL/TLS Based on Quantum Key Distribution," *Proceedings of the International Conference on Computer and Communication Engineering* (ICCCE08), Vol. I, pp. 919-922, Malaysia, May 13-15, 2008.

[10] Maxime Buser, Rafael Dowsley, Muhammed Esgin, Clémentine Gritti, Shabnam Kasra Kermanshahi, Veronika Kuchta, Jason Legrow, et al, "A survey on exotic signatures for post-quantum blockchain: Challenges and research directions," *ACM Computing Surveys*, Vol. 55, No. 12, Article No. 251, 2023, pp. 1 – 32, doi:10.1145/3572771

[11] Gorjan Alagic, Daniel Apon, David Cooper, Quynh Dang, Thinh Dang, John Kelsey, Jacob Lichtinger, et al., "Status Report on the Third Round of the NIST Post-Quantum Cryptography Standardization Process," *National Institute of Standards and Technology* Interagency or Internal Report, NIST IR 8413-upd1, 102 pages, July 2022, doi: 10.6028/NIST.IR.8413-upd1.

[12] A. Holcomb, G. Pereira, B. Das, and M. Mosca, "PQFabric: A Permissioned Blockchain Secure from Both Classical and Quantum Attacks," *IEEE International Conference on Blockchain and Cryptocurrency* (ICBC), Sydney, Australia, 2021, pp. 1-9, doi: 10.1109/ICBC51069.2021.9461070.

[13] Peijun Zhang, Lianhai Wang, Wei Wang, Kunlun Fu, and Jinpeng Wang, "A Blockchain System Based on Quantum-Resistant Digital Signature," *Security and Communication Networks*, Vol. 2021, Article ID: 6671648, 13 pages, doi: 10.1155/2021/6671648.

[14] Cosimo Michelagnoli, "Quantum-resistant Blockchain: Introduction of Post-Quantum Cryptography in Hyperledger Fabric," Master Degree Thesis, Computer Engineering, Politecnico Di Torino, 2023.

[15] Bakhtiyor Yokubov, "Post-Quantum Blockchain for Internet of Things Domain," Doctor of Philosophy Thesis, Department of Electronic and Electrical Engineering, College of Engineering, Design and Physical Sciences, Brunel University London, United Kingdom, October 2023.

[16] Sufyan Al-Janabi, "Towards a Quantum-Resistant Blockchain based on QKD," *The 5<sup>th</sup> International Conferences on Communications Engineering and Computer Science* (CIC-COCOS'2024), Cihan University-Erbil, KRG, Iraq, April 2024, pp. 114-121, DOI: http://doi.org/10.24086/cocos2024/paper.1528.